\documentstyle[11pt,amssymb,amstex]{amsart}

\numberwithin{equation}{section}

\newcommand{\bea}{\begin{eqnarray}}
\newcommand{\eea}{\end{eqnarray}}
\newcommand{\ba}{\begin{array}}
\newcommand{\ea}{\end{array}}
\newcommand{\edc}{\end{document}}
\newcommand{\bc}{\begin{center}}
\newcommand{\ec}{\end{center}}
\newcommand{\be}{\begin{equation}}
\newcommand{\ee}{\end{equation}}

\def\bc{{\mathbb C}}

\def\bn{{\mathbb N}}

\def\a{\alpha}

\def\e{\epsilon}
\def\l{\lambda} 

\def\m{\mu}
\def\p{\psi}
\def\n{\nu}
\def\r{\rho}
\def\s{\sigma} \def\Z{\Sigma}

\def\f{\varphi}

\def\w{\omega}

\def\e{{\bf 1}\!\!{\rm I}}
\def\o{\otimes}
\newcommand{\ol}{\overline}

\newtheorem{thm}{Theorem}[section]
\newtheorem{lem}[thm]{Lemma}

\newtheorem{prop}[thm]{Proposition}

\theoremstyle{remark}
\newtheorem{rem}{Remark}[section]

\begin{document}

\title[ON ENTROPY TRANSMISSION]
{ON ENTROPY TRANSMISSION FOR QUANTUM CHANNELS}


\author{Nasir Ganikhodjaev}
\address{Nasir Ganikhodjaev\\
Faculty of Science\\
IIUM, P.O.Box, 141, 25710, Kuantan, Pahang, Malaysia } \email{{\tt
nasirgani@@yandex.ru}}

\author{Farrukh Mukhamedov}
\address{Farrukh Mukhamedov\\
Departamento de Fisica,
Universidade de Aveiro\\
Campus Universitario de Santiago\\
3810-193 Aveiro, Portugal} \email{{\tt far75m@@yandex.ru}, {\tt
farruh@@fis.ua.pt}}

\begin{abstract}
In this paper a notion of entropy transmission of quantum channels
is introduced as a natural extension of Ohya's entropy. Here by
quantum channel is meant unital completely positive mappings (ucp)
of $B(H)$ into itself, where $H$ is an infinite dimensional
Hilbert space. Using a representation theorem of ucp mapping we
associate to every ucp map a uniquely determined state, and prove
that entropy of ucp map is less then Ohya's entropy of the
associated state.
 \vskip 0.3cm \noindent {\it
Mathematics Subject Classification}: 94A17, 81Q99, 81P68, 94A15.\\
{\it Key words}: entropy, unital completely positive map, Ohya's
entropy.
\end{abstract}

\maketitle

\section{Introduction}

The concept of state in a physical system is a powerful weapon to
study the dynamical behavior of that system. Since von Neumann
introduced a quantum mechanical entropy of a state, many
physicists have applied it in several dynamical systems and
studied its general properties (see \cite{L}, \cite{W}). For
several reasons this entropy had been extended to $C^*$-dynamical
systems by \cite{A},\cite{AL},\cite{L},\cite{U}. In \cite{O1} the
entropy of a state in quantum systems within $C^*$-algebraic
framework was introduced and studied. Note that the mechanism of
transmission is expressed by a so-called channel between input and
output receivers. In the quantum information theory channels play
an important role (see for example \cite{IKO},\cite{O2,O3,
O4},\cite{OP}). In \cite{OP},\cite{OPW} the quantum mutual entropy
and the quantum capacity were studied. In the algebraic framework
a quantum channel can be expressed by so-called completely
positive mapping from a $C^*$-algebra $M$ into another a
$C^*$-algebra $N$. In the present paper we introduce a natural
definition of entropy of unital completely positive (ucp) maps of
$B(H)$. We show that this entropy coincides with Ohya's entropy of
a state (see \cite{O1}) when we take a state instead of ucp map.
Note that in \cite{AOW},\cite{KOW} dynamical entropy of ucp maps
were introduced and studied, which differs from ours. Since here
introduced entropy in some sense, as well as, is an extension of
von Neumann's quantum mechanical entropy  to ucp maps. On the
other hand, roughly speaking, each ucp map $T$ can be represented
as a convex combination of extremal ucp maps $T_k$, and hence,
information is going through a channel $T$ depends on information
going through channels $T_k$. The introduced entropy deals with
uncertainties coming in this way and measures the amount of chaos
within mixture of quantum channels.

\section{Preliminaries}

Let $B(H)$ be the set of all linear bounded operators defined on a
separable Hilbert space  $H$. An element  $x\in B(H)$ is called
{\it positive } if there is an element $y\in B(H)$ such that
$x=y^*y$. The set of all positive elements of $B(H)$ we denote by
$B(H)_+$. A linear functional $\w$ on $B(H)$ is said to be  {\it a
state} if $\w(x)\geq 0$ for all $x\in B(H)_+$ and
 $\w(\e)=1$, here $\e$ stands the identity operator of $H$.
A state  $\w$ is called  {\it faithful} if $\w(x^*x)=0$ implies
$x=0$. A state  $\w$ is called
 {\it trace} if the equality  $\w(xy)=\w(yx)$ is valid for all $x,y\in B(H)$.
A linear  functional $f$ is   called {\it normal} if for every
bounded increasing net $\{x_{\a}\}$ of positive elements of $B(H)$
the equality $ \sup_{\a}f(x_{\a})=f(\sup_{\a}x_{\a})$ is valid. By
$B(H)_*$ we denote the set all  linear normal functionals on
$B(H)$. By $S$ the set of normal states on $B(H)$ is denoted.

The set  of linear continuous (in norm ) maps of $B(H)$ into
itself is denoted by $BB(H)$. On $BB(H)$ we define a {\it weak
topology} by seminorms
\begin{equation}\label{semi}
p_{\f,x}(T)=|\f(Tx)|, \ \ x\in B(H), \f\in B(H)_*.
\end{equation}
A norm of $T\in BB(H)$ is defined as usual by
$$
\|T\|=\sup_{x\in B(H), \|x\|=1}\|T(x)\|.
$$
Denote
$$
B_1=\{T\in BB(H): \|T\|=1\}.
$$

In \cite{M}  the following theorem was proved.

\begin{thm}\label{weakc} The set $B_1$ is weak compact. \end{thm}

Recall that a linear map $T\in BB(H)$ is said to be {\it
completely positive} if for any two collections $a_1,\cdots,a_n\in
B(H)$, $b_1,\cdots,b_n\in B(H)$ the following relation holds
\begin{equation}\label{CP}
\sum_{i,j=1}^nb_i^*T(a_i^*a_j)b_j\geq 0.
\end{equation}
A completely positive map $T:B(H)\to B(H)$ with $T\e=\e$ is called
{\it unital completely positive (ucp)} map. The set of all ucp
maps defined on $B(H)$ we denote by $\Z(B(H))$. For a ucp map $T$
we have $\|T\|=\|T(\e)\|=1$, therefore $\Z(B(H))\subset B_1$.

\begin{prop}\label{weakc1} The set $\Z(B(H))$ is weak convex compact.
\end{prop}

\begin{pf}  Let a net  $\{T_{\n}\}\subset \Z(B(H))$
weakly converge to an operator $T$. This means that for any state
$\f\in S$ we have
\begin{equation}\label{1}
\f(T(x))= \lim_{\n\to\infty}\f(T_{\n}(x)) \ \qquad \forall x\in
B(H).
\end{equation}

Now show that $T$ is ucp map. From \eqref{1} one can see that
$T\e=\e$. Since every $T_\n$ is ucp map, so for them \eqref{CP}
holds, i.e.
\begin{equation}\label{CP1}
\sum_{i,j=1}^n\f(b_i^*T_\n(a_i^*a_j)b_j)\geq 0.
\end{equation}
Now passing to limit $\n\to\infty$ from both sides of \eqref{CP1}
we obtain that $T\in\Z(B(H))$. Therefore, $\Z(B(H))$ is a closed
subset of $B_1$. Now Theorem \ref{weakc} implies the assertion.
\end{pf}

Further $extr(\Z)$ denotes the set of all extremal points of a set
$\Z:=\Z(B(H))$. According to Proposition \ref{weakc1} and Krein -
Milman Theorem the set $extr(\Z)$ is non empty. Note that in
\cite{P} certain properties of the set  $extr(\Z)$ were studied.

\section{An entropy of ucp maps}

According to the compactness of $\Z$ we can apply the theory of
decompositions of Choquet (see \cite{BR}).

According to that theory \cite{BR} for every operator $T\in \Z$
there is a probability measure $\m$  on $extr(\Z)$ with a
barycenter on $T$ such that
\begin{equation}\label{T1}
T=\int_{extr(\Z)}hd\m(h)
\end{equation}
If the measure $\m$ is atomic, i.e. $\mu=\{\l_n\}$, $\l_n\geq 0$
and $\sum_{k=1}^{\infty}\l_k=1$ then \eqref{T1} has a form
\begin{equation}\label{T2}
T=\sum_{n=1}^{\infty}\l_nT_n, \ \ T_n\in extr(\Z).
\end{equation}
 An {\it  entropy} of a ucp map $T$ is defined by
\begin{equation}\label{HT}
H(T)=\inf\bigg\{-\sum_{n=1}^{\infty}\l_n\ln\l_n\bigg\},
\end{equation}
here  $\inf$ is taken over for all possible discrete
decompositions of $T$ because the measure $\mu$ is not always
unique. If $\mu$ is not atomic then $H(T)$ is defined to be
infinite, i.e. $H(T)=\infty.$

It is easy to see that  $H(T)=0$  if and only if  $T\in extr(\Z)$.

Denote
\begin{equation*}
\Z_f=\{T\in \Z : H(T)<\infty\}.
\end{equation*}

\begin{thm}\label{Zf}  The weak closure of $\Z_f$ is the set $\Z$.
\end{thm}

The proof immediately follows from Proposition \ref{weakc1} and
the Krein-Milman Theorem.

For a given state  $\f\in S$ we define an ucp map $T_{\f}$ by
\begin{equation*}
T_{\f}x=\f(x)\e, \ \ \ x\in B(H).
\end{equation*}

\begin{lem}\label{ext} If  $\f\in extr(S)$ then $T_{\f}\in extr(\Z(B(H)))$.
\end{lem}

\begin{pf} Let us assume that the equality $T_{\f}=\l
T_1+(1-\l)T_2$ is valid, here $T_1,T_2\in \Z(B(H))$  and $\l\in
(0,1)$. Whence for every state $\p\in S$ we find
\begin{equation*}
\f(x)=\l\p(T_1x)+(1-\l)\p(T_2x), \ \ \ x\in B(H).
\end{equation*}
According to the extremity of the state $\f$ we get
$\p(T_ix)=\f(x), \ i=1,2$. From the arbitrariness of  $\p$ we
conclude that $T_ix=\f(x)\e$. Thus $T_{\f}\in extr(\Z(B(H)))$. The
lemma is proved.
\end{pf}

We recall that  an Ohya's-entropy  of a state  $\f\in S$
introduced in \cite{O1} and denoted by $h(\f)$.  Given a state
$\f\in S$ there is a positive operator $\theta$ such that
$\f(x)=tr(\theta x)$, $x\in B(H)$. According to the spectral
resolution theorem the operator $\theta$ can be expressed by
$$
\theta=\sum_n\l_n\phi_n
$$
where $\l_n$ are eigenvalues of $ \theta$, and $\phi_n$ are
projections to one dimensional subspaces generated by mutually
orthogonal eigenvectors   associated with $\l_n$, i.e.
$$
\phi_n(\xi)=(\xi,\phi_n)\phi_n, \ \ \xi\in H, \ n\in\bn.
$$
Then {\it Ohya's entropy} of a state $\f$ is defined by
\begin{equation}\label{Oh}
h(\f)=\inf\bigg\{-\sum_{n=1}^{\infty}\l_n\ln\l_n\bigg\},
\end{equation} here $\inf$ is taken over for all possible discrete
decompositions of $\theta$.

According to Lemma \ref{ext} with \eqref{HT},\eqref{Oh} we have
$$
H(T_{\f})=h(\f).
$$

This shows that the introduced entropy $H(T)$ is a generalization
of Ohya's one.

Recall that a ucp map $T\in Z$ is called {\it normal} if for every
bounded increasing net of positive elements $\{x_{\s}\}$ of $B(H)$
the equality
$$
T(\sup_{\a}x_{\a})=\sup_{\a}T(x_{\a}).
$$
is valid. This definition is equivalent to that  an operator $T$
is continuous in $\s(B(H),B(H)_*)$ - weak topology.

Further by $tr$ we will denote a normalized trace on  $B(H)$.
Consider a set
$$
{\cal T}_0=\{x\in B(H)\o B(H) : tr\o tr(|x|)<\infty \}
$$
and denote  $\|x\|_1=tr\o tr (|x|), \ x\in {\cal T}_0$, which is a
norm. By  ${\cal T}(H\o H)$ we will denote the norm $\|\cdot\|_1$
closure of ${\cal T}_0$, and by $tr_H$ a conditional trace from
$B(H)\o B(H)$ onto $B(H)$ defined on the elements of kind $x\o y$
by
$$
tr_H(x\o y)=tr(y)x, \ \ x,y\in B(H).
$$

Now we formulate some auxiliary facts.

\begin{lem}\label{p0} If the following equality holds
\begin{equation}\label{p01}
tr_H(p(\e\o x))=0,
\end{equation}
for any  $x\in B(H)$, then  $p=0$.
\end{lem}

\begin{pf} It follows from  \eqref{p01} that  for
every $y\in B(H)$ the equality holds
\begin{eqnarray*}
0&=&tr((y\o\e) tr_H(p(\e\o x))=\\
&=&tr\o tr(p(y\o x))
\end{eqnarray*}
Hence from arbitrariness of  $x$ and $y$ we find $tr\o tr(pu)=0$
for any $u\in B(H)\o B(H)$, which implies that $p=0$.
\end{pf}

Let $\{\f_n\}$ be a complete orthonormal basis in the Hilbert
space $H$. One can see that a system $\{\f_m\o\f_n\}$ forms a
complete orthonormal basis for $H\o H$.

Denote
\begin{equation}\label{H0}
H_0=\bigg\{\w=\sum_{i,j=1}^{n}a_ib_j\f_i\o\f_j : \
\{a_i\}_{i=1}^n,\{b_i\}_{i=1}^n\subset\bc, \ n\in \bn\bigg\}.
\end{equation}

\begin{lem}\label{H01} The set $H_0$ is a dense subspace of $H\o H$.
\end{lem}

\begin{pf} It is know \cite{T} that the following set
\begin{equation*}\label{H02}
H_1=\bigg\{\sum_{i,j=1}^{n,m}a_ic_j\f_i\o\f_j : \
\{a_i\}_{i=1}^n,\{c_i\}_{i=1}^m\subset\bc, \ n,m\in \bn\bigg\}
\end{equation*}
is dense in $H\o H$. Therefore, it is enough to show $H_0= H_1$.
Due to obvious inclusion $H_0\subset H_1$, we have to show
$H_0\supset H_1$. Take any $v\in H_1$, i.e.
$$
v=\sum_{i,j=1}^{n,m}a_ic_j\f_i\o\f_j,
$$
where $\{a_i\}_{i=1}^n,\{c_j\}_{j=1}^m\subset\bc$. Without loss of
generality we may assume that $n\geq m$. Let us put
\begin{equation*}
b_k= \left\{
\begin{array}{ll}
c_k, \ \ \ \textrm{if} \ \ k\leq m,\\
0, \ \ \ \ \ \textrm{if} \ \ m<k\leq n.
\end{array}
\right.
\end{equation*}
Then one can see that
$$
v=\sum_{i,j=1}^{n}a_ib_j\f_i\o\f_j,
$$
hence $v\in H_0$. This proves the assertion.
\end{pf}

\begin{thm}\label{rep}  Let  $T\in B_1$. The following assertions are equivalent:
\begin{enumerate}
\item[(i)] $T$ is a normal ucp map;\\
\item[(ii)]There exists a unique positive operator $p\in {\cal
T}(H\o H)$ with $tr_H(p)=\e$ such that
$$
Tx=tr_{H}(p(\e\o x)), \ \ \ x\in B(H).
$$
\end{enumerate}
\end{thm}

\begin{pf} The implication (ii)$\Rightarrow$ (i) is obvious. Therefore, consider
(i)$\Rightarrow $(ii).
 Define  $e_{ij}:H\to H$ by
\begin{equation}\label{eij}
e_{ij}(\eta)=(\eta,\f_j)\f_i, \ \  \eta\in H, \ i,j\in \bn.
\end{equation}
It is clear that  $e_{ij}(\f_k)=\delta_{kj}\f_i$ and  $e_{ii}$ is
a projection to one dimensional subspace of $H$ generated by the
vector $\f_i$, where $\delta_{kj}$ stands for the symbol
Kornecker. Moreover $\{e_{ij}\}$ forms a basis for $B(H)$.

Now define
\begin{equation*}
p_{(ij)(kl)}=(T(e_{ij})\f_k,\f_l), \ \ \ i,j,k,l\in\bn.
\end{equation*}

Recall some properties of $p_{(ij)(kl)}$ (see \cite{S}, Chap. 4):

\begin{enumerate}
\item[({\bf A})]  for every $n\in\bn$ and any collection of
numbers $\{a_k\}_{k=1}^n, \{b_k\}_{k=1}^n$ the following
inequality holds
$$
\sum_{i,j,k,l=1}^{n}a_i\ol{a_j}b_k\ol{b_l}p_{(ij)(kl)}\geq 0;
$$

\item[({\bf B})] for every $i,j,k,l\in\bn$
$$
p_{(ij)(kl)}=p_{(ji)(lk)};
$$

\item[({\bf C})] For every $k,l\in\bn$
$$
\sum_{i=1}^{\infty}p_{(ii)(kl)}=\delta_{k,l}.
$$
\end{enumerate}

Define an operator  $p$ by
\begin{equation}\label{p}
p=\sum_{(ij)(kl)}p_{(ij)(kl)}e_{kl}\o e_{ij},
\end{equation}
here convergence by the norm $\|\cdot \|_1$.

Now we calculate the conditional trace of $p$:
\begin{eqnarray*}
tr_H(p)&=&tr_H\bigg(\sum_{(ij)(kl)}p_{(ij)(kl)}e_{kl}\o e_{ij}\bigg)\\
&=&\sum_{(ij)(kl)}p_{(ij)(kl)}tr(e_{ij}e_{kl}) \\
&=&\sum_{(ij)(kl)}p_{(ij)(kl)}\delta_{ij}e_{kl} \\
&=&\sum_{k}e_{kk}=\e,
\end{eqnarray*}
here we have just used the property (C) of $p_{(ij)(kl)}$.

Using \eqref{eij} and \eqref{p} consider an action of  $p$ to the
element $\f_q\o\f_m$:
\begin{eqnarray}\label{p-act}
p(\f_q\o\f_m)&=&\sum_{(ij)(kl)}p_{(ij)(kl)}e_{kl}(\f_q)\o e_{ij}(\f_m) \nonumber \\
&=&\sum_{(ij)(kl)}p_{(ij)(kl)}\delta_{lq}\f_k\o \delta_{jm}\f_i \nonumber \\
&=&\sum_{i,k}p_{(im)(kq)}\f_k\o \f_i  \end{eqnarray}

Now show that $p$ is positive. Indeed, take any $\w\in H_0$ (see
\eqref{H0}) , i.e.
$$
\w=\sum_{i,j=1}^{n}a_ib_j\f_i\o\f_j,
$$
where $\{a_i\}_{i=1}^n,\{b_i\}_{i=1}^n\subset\bc$,$n\in\bn$.

Thanks to the positivity of operator $T$, the property (A) and
\eqref{p-act}  we have
\begin{eqnarray*}
(p(\w),\w)&=&
\sum_{i,j=1}^{n}\sum_{k,l=1}^{n}a_ib_j\ol{a_k}\ol{b_l}
(p(\f_i\o\f_j),\f_k\o\f_l) \\
&=&\sum_{i,j,k,l=1}^{n}a_i\ol{a_k}b_j\ol{b_l}\sum_{r,s}p_{(rj)(si)}
(\f_s\o\f_r),\f_k\o\f_l) \\
&=&\sum_{i,j,k,l=1}^{n}a_i\ol{a_k}b_j\ol{b_l}p_{(lj)(ki)}\geq 0.
\end{eqnarray*} From the density of $H_0$ in $H\o H$ (see Lemma \ref{H01})
we get the positivity of $p$. Thus $p\in {\cal T}(H\o H)$.

Consider
\begin{eqnarray*}
p(\e\o e_{ij})&=&\sum_{(kl)(rs)}p_{(kl)(rs)}e_{rs}\o (e_{kl}\cdot e_{ij})\\
&=&\sum_{k,s}p_{(ki)(rs)}e_{rs}\o e_{kj}.
\end{eqnarray*}
 Then we
obtain
\begin{eqnarray*}
tr_H(p(\e\o e_{ij}))
&=&\sum_{k,s}p_{(ki)(rs)}tr(e_{kj})e_{rs}\\
&=&\sum_{k,l}p_{(ij)(rs)}e_{rs}=T(e_{ij}).
\end{eqnarray*}

Using continuity of $T$  for every  $x\in B(H)$ we get
$$
Tx=tr_H(p(\e\o x)).
$$

Let us show the uniqueness of $p$. Assume that there is another
operator $q$ such that $Tx=tr_H(q(\e\o x))$. Then we have $
tr_H((p-q)(\e\o x))=0$ for every  $x\in B(H)$. According to Lemma
\ref{p0} we obtain $p=q$. This completes the proof. \end{pf}

\begin{rem} The proved Theorem generalizes a result of
\cite{FA}, where similar result has been obtained over
finite-dimensional Hilbert spaces.
\end{rem}

The operator $p$ in Theorem \ref{rep} is called a {\it
representative operator} for $T$, and denoted by $\r_T$. For each
ucp map we can associate a state $\f_T$ on $B(H)\o B(H)$ defined
by
$$
\f_T(x)=tr\o tr(\r_Tx), \ \ \ x\in B(H)\o B(H).
$$

\begin{thm}\label{hH1} Let T be a ucp map on $B(H)$. Then the
following inequality holds
\begin{equation}\label{hH2}
H(T)\leq h(\f_T)=-tr\o tr(\r_T\ln\r_T). \end{equation}
\end{thm}

\begin{pf}  Let us decompose the operator $\r_T$ as follows
\begin{equation}\label{hH3}
\r_T=\sum_k\l_k\r_k,
\end{equation}
where $\l_k\geq 0$, $\sum_k\l_k=1$ and $tr_H(\r_k)=\e$. Then (see
\eqref{HT})
\begin{equation*}\label{hH1}
H(T)=\inf\bigg\{-\sum_{k}\l_k\ln\l_k\bigg\}.
\end{equation*}
According to the spectral resolution Theorem  each operator $\r_k$
can be decomposed as follows
\begin{equation}\label{hH4}
\r_k=\sum_n\mu_{k,n} e^{(k)}_n.
\end{equation}
Keeping in mind \eqref{hH4} we can rewrite \eqref{hH3} by
\begin{equation*}
\r_T=\sum_{k,n}\l_k\m_{k,n} e^{(k)}_n.
\end{equation*}

Thanks to $tr\otimes tr(\r_k)=1$ we get $\sum_n\m_{k,n}=1$. Hence,
using the definition of Ohya's entropy (see \eqref{Oh}) one gets
\begin{eqnarray*}
h(\f_T)&=&\inf\bigg\{-\sum_{k,n}\l_k\m_{k,n}\ln(\l_k\m_{k,n})\bigg\}\\
&=&\inf\bigg\{-\sum_{k,n}\l_k\m_{k,n}\ln\l_k-\sum_{k,n}\l_k\m_{k,n}\ln\m_{k,n}\bigg\}\\
&=& \inf\bigg\{-\sum_{k}\l_k\ln\l_k-\sum_{k,n}\l_k\m_{k,n}\ln\m_{k,n}\bigg\}\\
&\geq & H(T).\end{eqnarray*}
\end{pf}

\begin{rem} In Theorem \ref{hH1} strict inequality can occur.
Indeed, consider the following example.
\end{rem}

{\sc Example.} Let $H=\bc^2$, then $B(H)=M_2(\bc)$, here
$M_2(\bc)$ is the algebra of $2\times 2$ matrices over complex
field $\bc$. By $e_{ij}$ we denote the matrix units of $M_2(\bc)$.
A commutative algebra generated by the matrix units $e_{11}$ and
$e_{22}$ is denoted by $CM_2$. We represent every element $ \left(
\ba{cc}
a_{11} & a_{12}\\
a_{21} & a_{22}\\
\ea \right) $ of $M_2(C)$ as a vector of $\bc^4$ by
$(a_{11},a_{22},a_{12},a_{21})$.

Now take $T\in\Z(CM_2)$ defined by
$$
T= \left( \ba{cccc}
p & 1-p & 0 & 0\\
q & 1-q & 0 & 0\\
0 & 0 & 0 & 0 \\
0 & 0 & 0 & 0\\
\ea \right),
$$
here $p,q\in (0,1)$. One can see that that extreme elements of
$\Z(CM_2)$ are the following ones
$$
T_1= \left( \ba{cccc}
1 & 0 & 0 & 0\\
0 & 1 & 0 & 0\\
0 & 0 & 0 & 0 \\
0 & 0 & 0 & 0\\
\ea \right), \ \ T_2= \left( \ba{cccc}
1 & 0 & 0 & 0\\
1 & 0 & 0 & 0\\
0 & 0 & 0 & 0 \\
0 & 0 & 0 & 0\\
\ea \right),
$$
$$
T_3= \left( \ba{cccc}
0 & 1 & 0 & 0\\
0 & 1 & 0 & 0\\
0 & 0 & 0 & 0 \\
0 & 0 & 0 & 0\\
\ea \right), \ \ T_4= \left( \ba{cccc}
0 & 1 & 0 & 0\\
1 & 0 & 0 & 0\\
0 & 0 & 0 & 0 \\
0 & 0 & 0 & 0\\
\ea \right).
$$

A decomposition of $T$ into a convex combination of $T_i$,
$i=\overline{1,4}$ is given by
\begin{eqnarray}\label{dec}
T=(1-q-d)T_1+(p+q+d-1)T_2+(1-p-d)T_3+dT_4,
\end{eqnarray}
here $d\in[\max\{0,1-p-q\},\min\{1-p,1-q\}]$.

Furthermore, we will assume that $p+q=1$. Then from \eqref{dec}
one finds
\begin{equation*}\label{Ht}
H(T)=\inf_{d\in[0,\min{p,q}]}\bigg\{-(p-d)\log(p-d)-(q-d)\log(q-d)-2d\log
d\bigg\}
\end{equation*}

Without loss of generality we may assume that $p\geq q$. Then
investigating extremum of a function
$$
F(x)=-(p-x)\log(p-x)-(q-x)\log(q-x)-2x\log x
$$
on $[0,q]$ we find that
\begin{equation}\label{Ht}
H(T)=-p\log p-q\log q.
\end{equation}

The representing operator $\rho_T$ can be written by
$$
\rho_T= \left( \ba{cccc}
p & 0 & 0 & 0\\
0 & 1-p & 0 & 0\\
0 & 0 & q & 0 \\
0 & 0 & 0 & 1-q\\
\ea \right).
$$
It is clear that
$$
d(\rho_T)=-(1-p)\log(1-p)-p\log p-q\log q-(1-q)\log(1-q).
$$
From this and \eqref{Ht} one can see that $H(T)<d(\rho_T)$.

\section{Conclusions}

In this paper a notion of entropy transmission of quantum channels
is introduced. Here by quantum channel we mean unital completely
positive (ucp) mappings of $B(H)$ into itself, where $H$ is an
infinite dimensional separable Hilbert space. We have shown that
this entropy coincides with Ohya's entropy of a state (see
\cite{O1}) if we take a state instead of ucp map. Therefore, the
introduced entropy is a natural extension of Ohya's one. Since
Ohya's entropy was a generalization of von Neoumann quantum
mechanical entropy, consequently the introduced entropy is also an
extension of the quantum mechanical one. besides, it measures the
amount of chaos within mixture of quantum channels. Therefore, it
differs from the dynamical entropy of ucp maps introduced in
\cite{AOW,KOW}. Furthermore, using a representation theorem of ucp
maps  we associate to every ucp map a uniquely defined state, and
prove that entropy of ucp map is less then Ohya's entropy of the
associated state. We hope that this entropy will have some
relations with the Holevo capacity of quantum channel (see
\cite{H}).

\section*{Acknowledgements} The second named author (F.M.) thanks the
FCT (Portugal) grant SFRH/ BPD/17419/2004. Useful suggestions of
an anonymous referee are appreciated.


\begin{thebibliography}{99}



\bibitem{AOW} L. Accardi, M. Ohya, N. Watanabe, Rep. Math. Phys. {\bf 38}, No.3, 457
(1996).

\bibitem{A} H. Araki, Publ. R.I.M.S. {\bf 11}, 809 (1976); {\bf 13}, 173 (1977).

\bibitem{AL} H.Araki, E.H.Lieb, Commun. Math. Phys. {\bf 18}, 160 (1970).

\bibitem{BR} O. Bratteli and D.Robimson, \textit{ Operator algebras and
quantum statistical mechanics I.} Springer-Verlag, Berlin, 1979.

\bibitem{H} A.S. Holevo, Probab. Theory Appl. {\bf 48}, 359, (2003)
 e-print quant-ph/0211170;

\bibitem{IKO} R.S. Ingarden, A. Kossakowski, M. Ohya,
\textit{Information dynamics and open systems. Classical and
quantum approach}. Fundamental Theories of Physics. {\bf 86},
Kluwer Academic Publishers, Dordrecht, 1997.

\bibitem{KOW} A. Kossakowski, M. Ohya, N. Watanabe,
Infin. Dimens. Anal. Quantum Probab. Relat. Top. {\bf 2}, No.2,
267 (1999).

\bibitem{L} G. Linbland,  Commun. Math.Phys. {\bf 33}, 205 (1973).

\bibitem{M} F.M. Mukhamedov, Methods Funct. Anal. Topology, {\bf 5}, 26 (1999).

\bibitem{O1} M. Ohya, Jour. Math. Anal. Appl. {\bf 100}, 222 (1984).

\bibitem{O2} M. Ohya, IEICE, J. {\bf 80} A, 1638 (1998).

\bibitem{O3} M. Ohya, IEEE Trans. Inf. Theory {\bf 29}, 770 (1983).

\bibitem{O4} M. Ohya, Open Sys. Information Dyn.  {\bf 6}, 69 (1999).
1
\bibitem{OP} M. Ohya, D.Petz, \textit{ Quantum Entropy and Its Use}, Sprinder-Verlag,
1993

\bibitem{OPW} M. Ohya, D. Petz, N. Watanabe, Probab. Math. Stat. {\bf 17}, 179
(1997).

\bibitem{P} K.R. Pathasarathy, Infin. Dimens. Anal.
Quantum Probab. Relat. Top. {\bf 1}, No.4, 599 (1998).

\bibitem{S} T.A. Sarymsakov, \textit{ An Introduction to Quantum Probability},
 Fan, Tashkent, 1985.

\bibitem{FA} A. Fujiwara, P.Algoet, Phys. Rev. A. {\bf 59}, 3290 (1999).

\bibitem{T} M. Takesaki, \textit{Theory of Operator algebras, I}, Springer,
Berlin--Heidelberg--New York, 1979.

\bibitem{U} H.Umegaki, Kodai Math. semin. reprs. {\bf 14}, 59 (1962).

\bibitem{W} A.Wehrl, Rev. Modern.Phys. {\bf 50}, 221 (1978).


\end{thebibliography}
\end{document}